10th U. S. National Combustion Meeting
Organized by the Eastern States Section of the Combustion Institute
April 23-26, 2017
College Park, Maryland

# A new jet-stirred reactor for chemical kinetics investigations

*Abbasali A. Davani[1]\*, Paul D. Ronney[1]*
*[1]University of Southern California*
*Los Angeles, California, United States*
*\*Corresponding Author Email: davanida@usc.edu*

**Abstract:** A novel jet-stirred reactor was designed to study combustion processes at low Damköhler number (Da, ratio of residence time to chemical time), i.e. chemical kinetics. In this new design, multiple impinging turbulent jets are used to stir the mixture. The goal of this work is to identify an optimal configuration of multiple pairs of impinging jets and outlet ports for as a Jet-Stirred Reactor (JSR) for chemical kinetics experiments. With this motivation, ANSYS-FLUENT computations using the RANS - Reynolds Stress Model were used to simulate mixing and reaction in such geometries and their performance was compared to classical JSR (4 Jets In Plus (+) Pattern (4JIPP) introduced by Matras & Villermaux 1973; Dagaut et al. 1986; etc.). Results showed that a configuration of 8 jets, each surrounded by a concentric annular outlet (CIAO), at the corners of an imaginary cube circumscribed by a spherical chamber provided far more uniform composition and temperature and thereby more nearly match the idealizations assumed in well-stirred reactor theory, even at values of Da higher than those accessible to other JSR experiments. Moreover, the CIAO design yielded inferred reaction rate constants that were much to the actual values than the classical JSR design.

## 1. Introduction

Well Stirred Reactor (WSR) is an ideal chamber in which mixture is perfectly mixed and homogeneous inside. The Jet-Stirred Reactor (JSR) [1] is a classical apparatus for studying the chemical kinetics of combustion reactions. In JSRs, several jets of combustible reactants are fed into a mixing chamber maintained at elevated temperatures and reacted. Analysis of the products of reaction as a function of reactant composition and residence time is used to infer the rates and pathways of reaction. The key assumption required to interpret the data is that mixing of reactants and products is complete and instantaneous, i.e. that the mixing time scale $\tau_M$ is much smaller than either the residence time scale $\tau_R$ or the chemical time scale $\tau_C$ so that no gradients of composition or temperature occur. The JSR has one compelling advantage for chemical kinetics experiments over shock tubes and laminar or turbulent tubular flow reactors which greatly simplifies data interpretation: the JSR is ideally zero-dimensional in both space and time whereas shock tubes are ideally zero-dimensional only in space and one-dimensional time and flow reactors are ideally zero-dimensional only in time and one-dimensional in space. Additionally, JSRs have an inherent advantage over tubular flow reactors: the residence time $\tau_R$ is almost completely decoupled from the mixing time $\tau_M$, the latter potentially being far smaller whereas in tubular flow reactors the mixing time and residence time both scale inversely with the mean velocity and thus cannot be decoupled. Another problem with tubular flow reactors is axial dispersion; that is, since the velocity profile is not purely plug-flow, reactants along the centerline will be convected downstream more rapidly than material near the tube wall, which renders the apparatus two-dimensional in space to some extent. Both issues can potentially be avoided with a well-designed JSR.

Of course there are challenges associated with JSR design and operation, the foremost of which are (1) ensuring rapid mixing of incoming reactants with the material already in the JSR, specifically to ensure $\tau_M \ll \tau_R$ and $\tau_M \ll \tau_C$ and (2) avoiding pre-reaction before the reactants enter the JSR. While many studies, e.g. [2], have addressed the consequences of an assumed level of imperfect mixing, the prediction and characterization of unmixedness has received very little attention. In fact, in a recent review paper Herbinet and Battin-Leclerc [3] state that, "…no work on this topic has been performed since 1986, and new work with up-to-date experimental techniques could bring valuable information to the subject." We shall show that existing reactors used for obtaining chemical kinetic data for combustion reactions may lead to significant discrepancies between inferred reaction rate constants and the actual values even for an extremely simple test problem. A new reactor design that promises to provide much smaller discrepancies and also addresses the pre-reaction issue is proposed in current study. It should be emphasized that the objective of this study is not to determine the reaction rates and mechanisms of a particular fuel, but rather to ascertain the viability of proposed apparatus as an improved apparatus for conducting chemical kinetics studies and determine the range of conditions for which data obtained in such an apparatus can be considered accurate and reliable. To overcome the limitations of current JSRs, we propose to employ a set of impinging jets of reactants with multiple outlet ports, all optimized for rapid mixing times compared to residence times.

While variety of geometries based on different icosahedral arrangements of inlet and outlet ports are investigated, for brevity here one which computations have shown is far superior to all others, namely a spherical chamber with novel arrangement of 8 jets at the corners of an imaginary cube (equivalently, a double-tetrahedral configuration), each with a concentric outlet port, which is called the Concentric Inlet And Outlet (CIAO) reactor (Figure 1) will be discussed. The inspiration for this type of geometry comes from the work of Hwang and Eaton [4] who used a similar arrangement of acoustically-forced "synthetic jets" to study particle settling in homogeneous, isotropic, zero-mean-flow turbulence. Proposed concentric jet arrangement of paired inlets and outlets is analogous to the synthetic jets but provides much higher turbulence intensities over a much larger fraction of the chamber volume. [5]

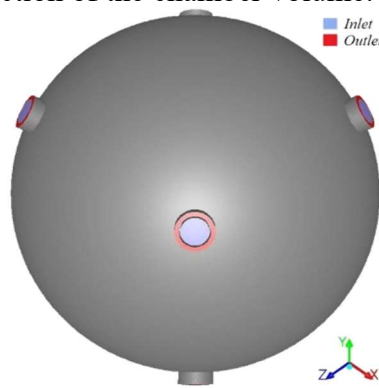

Figure 1. Proposed Concentric Inlet And Outlet (CIAO) reactor

## 2. Numerical Model and Feasibility Study

Aforementioned reactors have been simulated using ANSYS-FLUENT and the Reynolds-Averaged Navier Stokes (RANS) model for simulating turbulent transport. Accuracy of the used models are investigated and compared to experiments and other numerical studies and details of comparison as well as full cold flow analysis could be found in [5,6]. Using these models has the advantages of (1) greatly shortening the design cycle time; (2) interfacing directly with CAD software and 3D printers; and (3) simplifying the sharing of input and output files with other research groups (as opposed to the use of in-house codes.)

To assess the viability of the CIAO configuration for JSR problems, a very simple test problem is employed, consisting of a one-step or two-step reaction between reactants A and B, in the 2-step case with an intermediate species C and final product D. Initially, the case with no heat release and no activation energy is studied and then effect of heat release with reaction activation energy is studied.

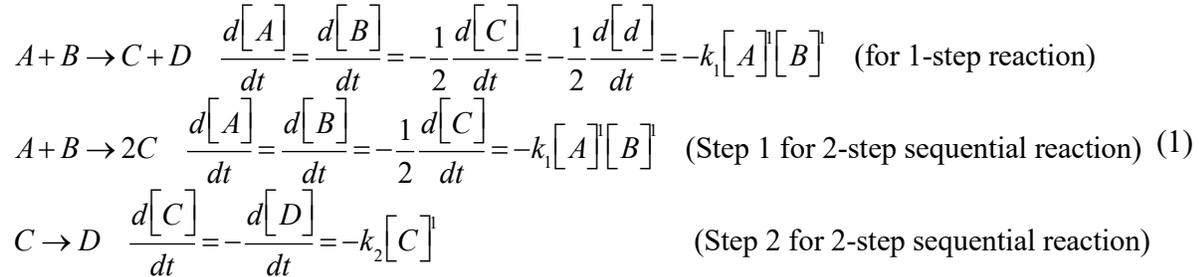

$$A + B \rightarrow C + D \quad \frac{d[A]}{dt} = \frac{d[B]}{dt} = -\frac{1}{2}\frac{d[C]}{dt} = -\frac{1}{2}\frac{d[d]}{dt} = -k_1[A][B] \quad \text{(for 1-step reaction)}$$

$$A + B \rightarrow 2C \quad \frac{d[A]}{dt} = \frac{d[B]}{dt} = -\frac{1}{2}\frac{d[C]}{dt} = -k_1[A][B] \quad \text{(Step 1 for 2-step sequential reaction)} \quad (1)$$

$$C \rightarrow D \quad \frac{d[C]}{dt} = -\frac{d[D]}{dt} = -k_2[C] \quad \text{(Step 2 for 2-step sequential reaction)}$$

If the mixture at the inlets is an equimolar mixture of A and B (i.e. their mole fractions $X_i$ are both 0.5) with no inert, the molar concentrations of A and B ([A] and [B] respectively, units moles/m³) are the same for all time since any reaction of A also causes a reaction of B. For this case, assuming perfect mixing, some algebra shows that the mole fractions $X_i$ in the reactor are

$$X_A = X_B = \frac{\sqrt{1+Da}-1}{Da}; \; X_C = X_D = \frac{1}{2} - \frac{\sqrt{1+Da_1}-1}{Da_1} \quad \text{(one-step)}$$

$$X_A = X_B = \frac{\sqrt{1+Da}-1}{Da}; \; X_C = \frac{2+Da-2\sqrt{1+Da}}{Da(1+\Delta Da)}; \; X_D = \Delta\frac{2+Da-2\sqrt{1+Da}}{(1+\Delta Da)} \quad \text{(two-step)} \quad (2)$$

$$\text{where } Da \equiv 4k_1[A]_o\frac{V}{\dot{V}}; \; \Delta \equiv \frac{k_2}{4k_1[A]_o}$$

where Da is a Damköhler number for step (1), $\Delta$ is the ratio of the Damköhler numbers for reaction step (2) to step (1), $V$ and $\dot{V}$ are the reactor volume and volume flow rate respectively (note $V/\dot{V}$ is the residence time $\tau_R$) and $[A]_o$ is the concentration of reactant A at the inlet. FLUENT / RANS modeling with this simple reaction scheme has been conducted for Dagaut et al.'s [1] reactor (Figure 2) and the CIAO reactor with the same V and $\tau_R$. In these computations, by intent no model of turbulence-chemistry interaction is used so that a direct comparison of the performance of the reactors can be made. Moreover, if a model of turbulence-chemistry interaction is needed the key assumption of JSRs – uniform composition, thus no variance in reaction – is already violated.

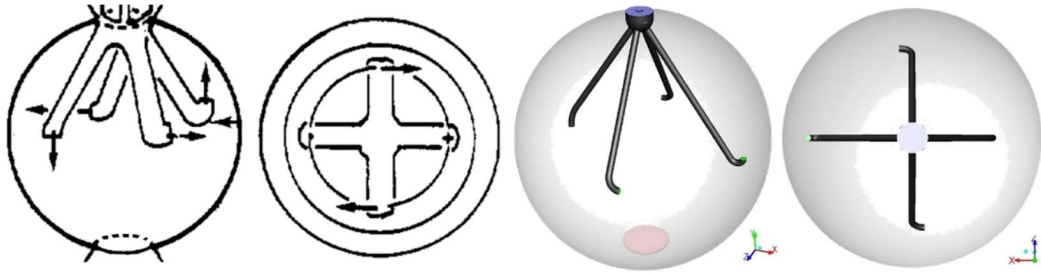

Figure 2. Dagaut *et al*. [1] reactor. Left 2 images: from the original paper; right 2 images: CAD model used in simulations. Overall diameter is 4 cm and the exit diameter of the 4 fuel jets is 0.1 cm.

## 3. Result and Discussion

### 3.1. Case I: No heat release and activation energy

Figure 3 (right) shows that the simulated volume-averaged values of $X_C$ in the CIAO reactor result in inferred values of $k_1$ that are much closer to the actual values than those obtained in the Dagaut-type reactor. These results suggest that even in this very simple chemical scenario, the non-uniformity of mixing in widely-used JSRs results in inferred reaction rate constants that are significantly different from the true values.

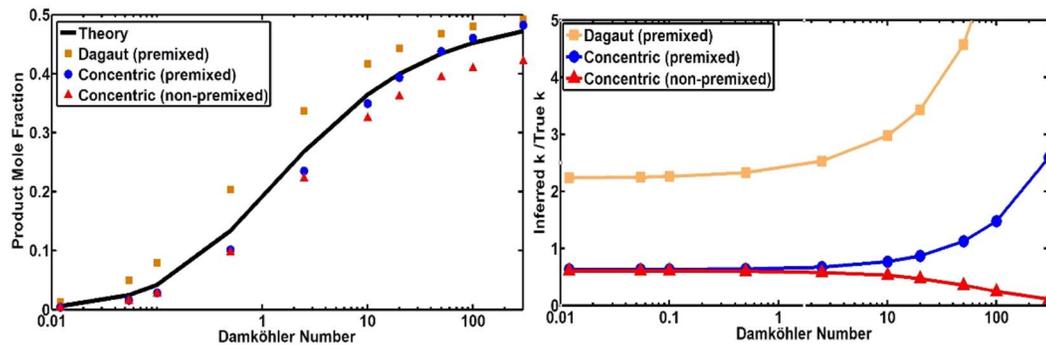

Figure 3. Left: Volume-averaged mole fraction of product C ($X_C$) as a function of Da for the test problem (Eq. (1)) with $\Delta = 0$ (i.e., single-step reaction) predicted by the FLUENT/RANS simulations for the Dagaut-type reactor and the concentric jet (CIAO) reactor along with comparison to the exact theory (Eq. (2)) assuming perfect mixing. Right: comparison of inferred (from Eq. (2)) rate constants using the volume-averaged product mole fraction for the Dagaut and CIAO reactors (relative to the actual prescribed k1) as a function of Da.

In order to facilitate comparisons between the Dagaut-type and CIAO reactors, the aforementioned simulations have of necessity employed premixed reactants with the presumption of no reaction before entering the chamber. This was necessary because the Dagaut-type reactor has one high mass flow inlet path for air (or other oxidant) and a different low-mass-flow, low-residence-time inlet path for fuel to minimize potential fuel pyrolysis. This is not directly compatible with the symmetrical inlets of the CIAO reactor, hence the use of premixed reactants for comparison. Of course, premixing of reactants is not a practical technique for real experiments because heated premixed reactants will begin to react before entering the "reactor." Consequently, simulations of the CIAO reactor were performed using non-premixed reactants (4 jets with reactant A, 4 jets with reactant B). Figure 3 shows that the results with non-premixed reactants are nearly the same as those of premixed reactants up to Da ≈ 10. Fundamentally this is possible because the CIAO reactor has effectively decoupled the residence time $\tau_R$ from the mixing time $\tau_M$, the latter being much smaller.

### 3.2.    Case II: Effects of heat release and activation energy

Most chemical reaction rates of relevance to combustion are far more sensitive to temperature (due to the Arrhenius term) than composition. Consequently, in JSRs, by intention the reactant concentrations are usually limited to values that are sufficiently small that the heat release does not cause a significant enough temperature rise to have a substantial effect on reaction rates. The drawback of low reactant concentrations (and thus low product concentrations) is that chemical analysis (e.g. gas chromatography and mass spectrometry) and optical diagnostics will be less accurate due to lower signal-to-noise ratios. For an Arrhenius temperature dependence, the ratio of reaction rate $\omega$ at the initial temperature $T_i$ to that at the adiabatic (complete reaction) temperature $T_f$ is given by

$$\frac{\omega\left(T_f\right)}{\omega\left(T_i\right)}=\exp\left(\beta\frac{T_f-T_i}{T_f}\right); \beta\equiv\frac{E}{RT_i} \tag{3}$$

where E is the effective activation energy for the reaction and R the gas constant. The issue lies in the fact that the non-dimensional activation energy $\beta$ is typically large for the reactions of relevance to combustion and thus $T_f - T_i$ must be small if the variation in $\omega$ due to temperature fluctuations is to be kept small. For example, for a typical overall activation energy of 30 kcal/mole, with $T_i = 900K$, to limit $\omega(T_f)/\omega(T_i) \leq 2$ requires $T_f \leq 1.043\ T_i = 939K$, which in turn for a lean propane-air mixture limits the fuel concentration to 1.5% of stoichiometric. The key question for modeling to address is, do the apparent advantages of the CIAO configuration compared to existing JSR designs still apply for systems with finite heat release and if so, what is the maximum heat release or more specifically $\beta(T_f-T_i)/T_f$ for which the well-stirred approximation can be employed? An analysis for a single-step reaction similar to the isothermal analysis above, including an enthalpy balance between the degree of reaction and the temperature with the reactor $T_R$ leads to a typical WSR relation of the form

$$Da^* \equiv Da\exp\left(\frac{-E}{RT_i}\right)=\left(\frac{T_R}{T_i}\right)^2\frac{\left(T_R-T_i\right)\left(T_f-T_i\right)}{\left(T_f-T_R\right)^2}\exp\left[\beta\left(\frac{1}{T_R}-\frac{1}{T_i}\right)\right] \tag{4}$$

where Da has been scaled so that the reaction rate with finite activation energy but zero heat release is the same as that with zero activation energy as in Eq. (1). The isothermal and non-isothermal predictions and comparison to computed results in the CIAO reactor with the conditions are shown in Figure 4. As a direct consequence of the scaling chosen, with heat release the mean reaction rate increases and thus more product is formed at a given Da. In the limit of low Da there is little heat release ($T_R \approx T_i$) and thus the results are the same for zero and finite activation energy. Figure 4 shows that the CIAO reactor is able to reproduce the effects of heat release and finite activation energy reasonably well, at least for the case $\omega(T_f)/\omega(T_i) \leq 2$.

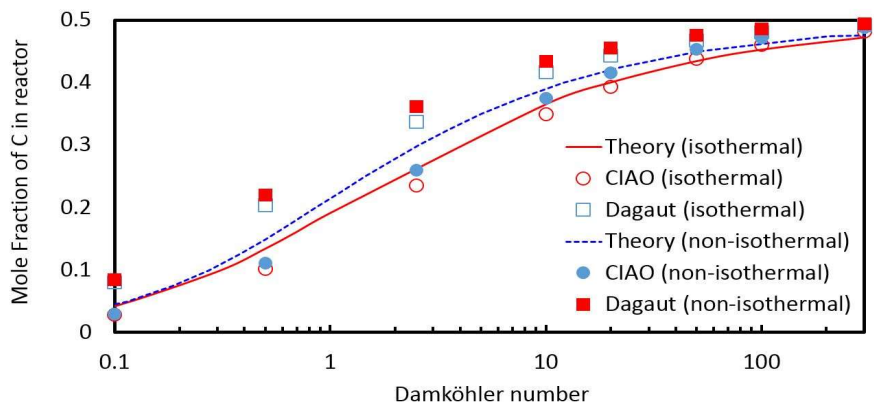

Figure 4. Comparison of computed product mole fractions ($X_C$) for single-step reaction in the CIAO to theoretical predictions for WSRs in under isothermal and non-isothermal conditions; for the latter case with E = 30 kcal/mole, $T_i$ = 900K, $T_f$ = 939K, resulting in $\omega(T_f)/\omega(T_i)$ = 2.

## 4. Conclusion

The proposed CIAO apparatus shows promising results in study of chemical kinetics. CFD computations for simple reaction models demonstrate that the apparatus provides an improved performance in study of chemical kinetics since even in very simple chemical scenarios, the non-uniformity of mixing in widely-used JSRs results in inferred reaction rate constants that are significantly different from the true values.

### Acknowledgments

Authors would like to appreciate Dr. Philippe Dagaut for providing useful information of their reactor and also for technical discussions This research was sponsored by U. S. National Science Foundation, Grant No. CBET-1236892 and the U. S. Air Force Office of Scientific Research, Grant No. FA9550-16-1-0197.